\documentclass[11pt]{article}
\usepackage{epsfig}
\begin{document}
%
%
%
%
\title
{Taking $N\to 0$ with $S$ matrices}

\author{Paul Fendley\\
Department of Physics\\
University of Virginia\\
Charlottesville, VA 22904-4714\\
{\tt fendley@virginia.edu}}

\maketitle

\centerline{\it Contribution to the proceedings of the}
\centerline{\it  NATO Advanced Research Workshop on
Statistical Field Theories, Como 18-23 June 2001}

\bigskip\bigskip
\begin{abstract}

Interesting physical results can be obtained from sigma models by
taking the number of fields $N$ to zero.  I discuss how one can make
sense of this limit by using exact $S$ matrix techniques. I review how
this can be done for the case of self-avoiding polymers, and speculate
on the application to the replica limit of disordered systems.

\end{abstract}

\section{Introduction}

Significant progress has been made on understanding the phase
structure of disordered models in two dimensions.  One key tool has
been in formulating the systems in terms of replica or supersymmetric
sigma models.  A sigma model is a field theory where the fields take
values on a manifold. Models can be classified according to their
symmetry \cite{Zirn}, so systems with very different physical origins
can still be described by the same sigma model.

The phase structures of many two-dimensional sigma models are fairly
well understood, so we know a great deal about the phases of the
corresponding disordered models (although the latter seems to exhibit
even richer behavior; see for example the discussion in
\cite{classD}).  The types of critical points possible have been
extensively studied \cite{nato}.
Much less progress has been made on understanding the non-trivial
critical points themselves. These points should have conformal
invariance and thus be described by conformal field theories. However,
only in a few cases \cite{GLR,GLL} has a non-trivial critical point
been definitively identified as a conformal field theory, and critical
exponents computed exactly.

The supersymmetric sigma models describing disordered systems are
well-defined. Unfortunately, many of the conventional theoretical
approaches to such problems are either extremely difficult or
impossible, because the bosonic part of the field manifold is
non-compact. The replica sigma models can be
formulated on compact manifolds. The (huge) disadvantage here is that
one must take the number of fields (the dimension of the field
manifold) to zero at the end of computation. This is at best
ill-defined outside of perturbation theory, and at worst completely wrong.

There is, however, one well-understood field theory where the number
of fields is taken to zero.  This is a self-avoiding random walk
(often known as a polymer) in two dimensions. The field-theory
description of this problem involves a field transforming as the
vector of $O(N)$, and at the end, one must send $N$ to zero. There are
formulations of the problem which are completely well-defined as a
function of $N$, and which are believed to be in the same universality
class. Namely, there is a lattice model defined for all $N$, not just
integer values \cite{nienhuis}. This continuum limit of this lattice
model for any $N$ can be mapped onto a field theory by Coulomb-gas
techniques. The mapping is not rigorous, but widely believed to give a
field theory in the correct universality class. Moreover, one can
extend these results even further by finding a description of the
polymer field theory in terms of exact $S$ matrices
\cite{Zpoly}. Again, this description is precisely defined for any
$N$, and one only need make the assumption that the models are in the
same universality class.

The aim of this contribution to the conference proceedings is to
explain the last paragraph in more detail, and to present some
speculations on the application to one particular type of system, the
metal-insulator transition critical point in the GSE class. This class
is interesting in that it is the two-dimensional version of one of the
original random matrix ensembles of Wigner and Dyson.  Since this
class of models is integrable for finite integer $N$
\cite{metheta,meTBA}, one might hope to be able to apply $S$ matrix
methods here as well.  One key problem I will discuss is the
singularity that appears here and in the polymer model as one
decreases the number of fields through $N=2$.

In section 2 I  review the arguments for the existence of a
metal-insulator transition in the GSE class. In
section 3, I review the solution of the 
dilute/dense transition for a self-avoiding
polymer. I show how this problem can be solved by applying $S$ matrix
methods. In section 4, I speculate on the applications of $S$ matrix
methods to the metal-insulator transition in the GSE class.

\section{The metal/insulator transition\\
in the GSE class replica sigma model}

The replica sigma model for the GSE class in two dimensions is a sigma model
where the field takes values in the manifold
$$
\frac{O(2P)}{O(P)\times O(P)}.
$$
For $P> 1$, the exact $S$ matrix for this sigma model has been found
\cite{metheta,meTBA}.
The action can be written in terms
of a $2P\times 2P$ matrix field $\Phi$:
\begin{equation}
S= \frac{1}{2g}
 \hbox{tr}\, \int d^2x\ \partial^\mu \Phi^\dagger \partial_\mu \Phi
\label{action}
\end{equation}
where $\Phi$ must be real, symmetric, orthogonal
and traceless.
The beta function for this model has been computed to at least three loops
\cite{hikami}. 
For the more general sigma model on the manifold $O(N)/O(N-P)\times O(P)$
the beta function at two loops is
\begin{equation}
 \beta(g) = -(N-2) g^2 -[2P(N-P) - N] g^3 
\label{beta}
\end{equation}
A number of important facts are apparent from this beta
function. First of all, there is a critical point at $g=0$. This is
the trivial fixed point, where the sigma model manifold effectively
becomes flat and the model reduces to free bosonic fields.  A crucial
consequence of this beta function is that for $N>2$ ($P>1$ for the
GSE case), is that the trivial fixed point is
unstable. At longer distance, the system flows away from the trivial
fixed point and the coupling $g$ increases. Another consequence of
(\ref{beta}) is that there is no evidence for a zero of $\beta(g)$ other
than the trivial fixed point for any $N\ge 0$.  This of course does
not prove that there is no such fixed point, only that it can not be
found by a perturbative expansion valid near the trivial fixed point.

The coupling $g$ increasing arbitrarily as the distance scale is
increased is a signal of Anderson localization, because $g\to\infty$
corresponds to the conductance decreasing to zero.  However, this is
{\bf not} what happens in the GSE class. The reason is that $\beta(g)$
for $g$ small changes sign when $N$ is decreased below $2$.  This
means that the $g=0$ trivial fixed point is {\bf stable} when $N<2$,
and so there is a {\it metallic phase} in the disordered system for
small enough $g$. For large enough $g$, one expects disorder to
dominate and localization to occur. This implies there is a
metal/insulator transition at some value $g_c$. Since no fixed point
appears in the beta function, the value of $g_c$ must be out of the
regime of validity of perturbation theory. These considerations lead
to the phase diagrams for the $O(2P)/O(P)\times O(P)$ sigma model at
$\theta=0$ displayed in the figure 1. The left one is valid for $N>2$,
while the right applies to $N<2$, and includes the replica limit of
the GSE class.

\begin{figure}
\centerline{\epsfxsize=4.5in\epsffile{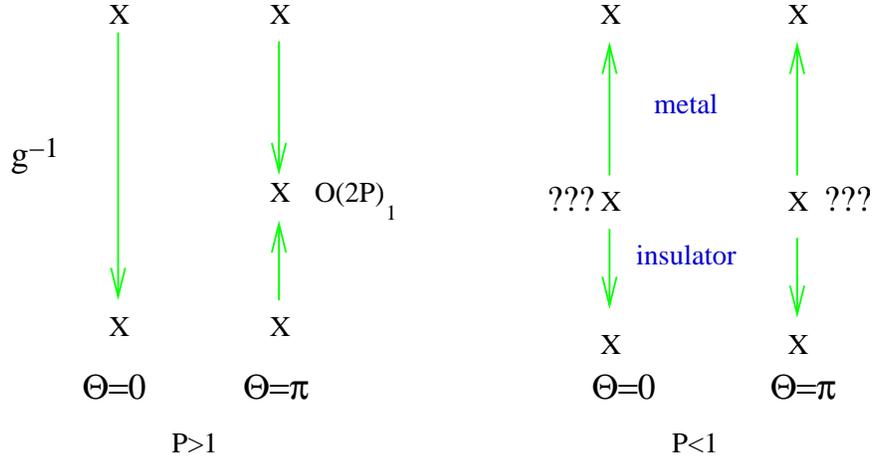}}
\caption{The proposed phase diagram for the $O(2P)/O(P)\times O(P)$
sigma model}
\end{figure}

A parameter called the $\theta$ angle appears in some sigma models.
This is crucial for example
in the replica approach to the plateau transition in the quantum Hall effect
\cite{pruisken}. The theta angle is
associated with field configurations called instantons. 
This is well known for $P=2$, where the
$O(4)/O(2)\times O(2)$ sigma model reduces to two copies of the
$O(3)/O(2)$ sigma model. In the $O(3)/O(2)$ model, the field takes values on
the sphere, and is believed to describe the continuum limit of
Heisenberg spin chains, with $\theta=0$ describing integer-spin
chains, while $\theta=\pi$ describes those with half-integer spin
\cite{Haldane}. The field theory with $\theta=0$ is gapped, and the
spectrum is a triplet under the global $O(3)$ symmetry, while when
$\theta=\pi$, the spectrum is gapless and forms doublets under the
global symmetry \cite{ZZtheta}. 
For the GSE class model for general $P$, 
the parameter $\theta$ can take just two values $\theta=0,\pi$.
For $P>2$, the behavior is similar to $P=2$:
the spectrum is gapped for $\theta=0$ and gapless for $\theta=\pi$
\cite{metheta}.

The sigma model at $\theta=\pi$ with $P>1$ therefore has a non-trivial
fixed point when $\theta=\pi$, as illustrated in the left half of
figure 1.  Since the value of $\theta$ does not affect perturbation
theory, the value of $g_c$ must be outside the region of validity of
perturbation theory. The question now is what happens for $P<1$.  For
$g$ small, perturbation theory is valid and the beta function
(\ref{beta}) is applicable to both $\theta=0$ and $\pi$. This means
that the flow for $P<1$ must be towards the trivial fixed point, at
least for $g$ small. One still expects that for $g$ large, the model
is in a disordered phase. The simplest possible phase diagram for
the $P<1$ sigma model is shown in the right half of figure 1. The 
non-trivial critical point now describes a metal-insulator transition.
Comparing the phase structure for $\theta=0$ and $\pi$, an
important distinction is that for the latter, the non-trivial critical
point is present even for $P>1$.

For $\theta=\pi$ and $P>1$, the conformal field theory describing the
critical point has been identified in \cite{metheta}. It is the
$O(2P)_1$ WZW theory, which essentially amounts to $P$ free Dirac
fermions. However, continuing this result to $P\to 0$ presents
immediate problems. The vanishing of the beta function is not the only
unusual thing happening at $N=2$. I will show in section 4 the exact
solution of \cite{metheta,meTBA} becomes singular there, making taking
the replica limit quite tricky.  This is fortunate, because it does
not seem likely that the metal-insulator transition in the GSE class
is described by free fermions.

\section{The $S$ matrix approach to polymers}

In the previous section I described how a ``metal'' phase appears in the
field theories with $N<2$. Even though the sigma
models are solvable for $N>2$, various pathologies happen to the
solution as $P$ is deformed to $1$. This sort of behavior happens in
all two-dimensional sigma models with $O(N)$ symmetry as $N$ is
decreased to $2$. This shows that one cannot simply continue the
results from $N>2$ to $N\to 0$.  Indeed, it even seems possible that
results in such models valid for $N>2$ are meaningless in the $N\to 0$
replica limit. However, one reliable prediction has been made already:
the beta function computed for $N>2$ implies a metallic phase in the
replica limit. The existence of a metallic phase has been fairly well
established for at least some models in class $D$, where the replica
sigma model is $O(2P)/U(P)$  (although results for this
model have numerous subtleties \cite{classD}).

The purpose of this section is to explain in detail a model with
$O(N)$ symmetry where the results for $N>2$ can be used to guide the
way to $N\to 0$. This is the field theory describing
self-avoiding random walks (often called polymers)
\cite{degennes}. The same pathologies that happen in the GSE class
sigma model as $N\to 2$ also happen here. Nevertheless, the non-trivial fixed
point when $N\to 0$ is fairly well understood \cite{nienhuis}. 
It is important to emphasize that is it not possible to naively continue the
results for $N>2$ to $N\to 0$. What I will argue is that results known
from solving the theory for $N>2$ imply certain criteria that the
$N\to 0$ theory must obey. These criteria allow one to solve the
theory for $N<2$.

The polymer critical point in two dimensions was first understood by
using Coulomb-gas methods. There is another approach available: the
$S$ matrix approach to integrable models. Here one works directly in
the continuum, but ends up reproducing and extending the results of
the Coulomb-gas approach \cite{Zpoly}.  These sigma models are defined
as classical two-dimensional field theories. In the $S$ matrix
approach, the model is treated as a $1+1$ dimensional quantum field
theory, so the rotational invariance of the classical model turns into
Lorentz invariance of the quantum model.  All the states of this
quantum theory can be described in terms of quasiparticles in
a Hilbert space. One can in principle (and
sometimes in practice) derive the masses and
multiplicities of these quasiparticles from the Bethe ansatz,
but it is usually easier to infer them from the symmetries of the
theory. The quasiparticle scattering matrix for an integrable model
must satisfy a number of extremely restrictive constraints. By
requiring that it satisfy these constraints, and agree
with the perturbative expansions and all the symmetries of the model
at hand, one can infer the $S$ matrix as well. 
Once the exact $S$ matrix
is known, essentially all thermodynamic quantities can
usually be derived.  In addition,
geometrical quantities like the scaling function for the
number of polymer configurations on a cylinder can be computed
\cite{FSpoly}, as well as some form factors \cite{CM}.

One nice feature of the polymer problem is that there is are several
precise ways to define the model for all $N$. One way is via an $O(N)$
vector model on the honeycomb lattice. A fixed-length $N$-component
vector ${\bf n}_i$ obeying ${\bf n}_i\cdot {\bf n}_i =1$
is placed on each site $i$ of the lattice. The
partition function of this two-dimensional classical model is taken to
be
\begin{equation}
Z = \int [d{\bf n}_i]
\prod_{<ij>}\left(1 + K {\bf n}_i\cdot {\bf n}_j\right)
\label{poly}
\end{equation}
where $\langle ij\rangle$ are nearest neighbors. By construction this model has a
global $O(N)$ symmetry. This choice of partition function should be in
the same universality class as the usual vector model with energy
$J\sum_{<ij>}{\bf n}_i\cdot {\bf n}_j$.  The reason for this choice is
that this model is equivalent to a model which not only can be defined
for any $N$, but which makes the relation to the polymer problem clear. By
performing a high-temperature expansion, the action (\ref{poly}) is
equivalent to a loop gas with action
\begin{equation}
 Z = \sum K^{n_{links}} N^{n_{loops}}
\label{loopgas}
\end{equation}
The sum is over all closed
self-avoiding and mutually-avoiding loop configurations on the
hexagonal lattice, with $n_{loops}$ the number of such loops, and
$n_{links}$ the number of links in these loops \cite{nienhuis}. 
Here $N$ appears in
the partition function only as a parameter, and so can now be taken to
be any value, not just the positive integers. In particular, when $N$
approaches zero, the partition function is dominated by the
configurations which have only a single loop. This loop is the
self-avoiding polymer.

This model has a critical point at
$K=K_c= (2+\sqrt{2-N})^{-1/2}$. This critical point is at real
values of $K$ only for $N\le 2$. Thus the
two-dimensional self-avoiding polymer $N\to 0$ has a non-trivial critical
point separating
the high-temperature ($K<K_c$)
``dilute'' phase from the low-temperature ($K>K_c$) ``dense'' phase.
In the dense phase, the preferred polymer configurations
cover essentially the entire plane. Exact critical exponents
describing the dilute-dense transition
were originally derived by applying Coulomb-gas methods to the action
(\ref{loopgas}) \cite{nienhuis}.

When $N$ is a positive integer, the continuum limit of the lattice
model (\ref{poly}) should be described by the ``$O(N)$'' non-linear
sigma model, where the field manifold is $O(N)/O(N-1)$. 
This $O(N)$-symmetric model is described by a real field $\vec{\phi}(x)$
with $N$ components taking values on the $(N-1)$-sphere, and so obeys
$\vec{\phi}\cdot\vec{\phi} =1$. It has action
\begin{equation}
S^{O(N)}=
\frac{1}{g}\int d^2 x\ \partial_\mu \vec{\phi} \cdot \partial_\mu \vec{\phi}
\label{onaction}
\end{equation}
Note that weak coupling in the sigma model ($g$ small) corresponds to
$K$ large in the lattice model. Hence the {\it dense} phase of the
polymer model is analogous to the metal phase of the disordered model,
while the dilute phase is analogous to the insulating phase.

The perturbative beta function (\ref{beta}) of this field theory for $N>2$
indicates that the trivial fixed point $g=0$ is unstable. There is no
evidence for a non-trivial low-energy fixed point, from the
beta function or otherwise (there are 
no instantons and hence no theta term for $N>3$).
Since the only fixed point is the
trivial one, it is not obvious how to obtain any information about
the dilute-dense transition by continuing the $N>2$ results for the
sigma model to $N\to
0$. Indeed, the $S$ matrix has a
singularity as $N\to 2$, and cannot be analytically continued
past. Nevertheless, there is an $S$ matrix describing the $O(N)$ model
for $N<2$ and in particular the dilute-dense transition as $N\to 0$.

The field theory (\ref{onaction}) is integrable for any integer
$N>2$. 
The simplest ansatz is that
there is a multiplet of $N$ massive particles
in the vector representation of $O(N)$ \cite{ZZ}.
A hint in favor of this ansatz is that
the fields in the action (\ref{onaction}) are in the vector
representation of $O(N)$. 
To check the ansatz thoroughly, one needs to compute
the $S$ matrix for these particles.
Because of the relativistic invariance of this
theory, it is convenient to use the rapidity variable $\beta_1$,
defined in terms of the quasiparticle mass, energy and momentum
as $E=m\cosh\beta_1$ $p=m\cosh\beta_1$.  
A two-particle
state is $|i_1(\beta_1) i_2(\beta_2)\rangle$, where
$i_1$ and $i_2$ label the member of the multiplet: $i_1,i_2 =1\dots
N$.  The two-particle $S$ matrix $S_{i_1 i_2}^{j_1 j_2}(\beta)$ is
the amplitude for scattering this initial state to the final state
$|j_1(\beta_1) j_2(\beta_2)\rangle$. The integrability requires that
the collision be completely elastic: the individual momenta do not
change. Lorentz invariance requires that the $S$ matrix element
depends only on the difference of the momenta:
$\beta=\beta_1-\beta_2$. The $O(N)$ invariance requires that
$S$ be of the form
\begin{equation}
S_{i_1 i_2}^{j_1 j_2}(\beta) = S_0(\beta) \delta_{i_1}^{j_1}
\delta_{i_2}^{j_2}  +  S_1(\beta) \delta_{i_1}^{j_2}
\delta_{i_2}^{j_1} +  S_2(\beta) \delta_{i_1i_2}
\delta^{j_1j_2} 
\label{onsmat}
\end{equation}
for some functions $S_0(\beta)$, $S_1(\beta)$ and $S_2(\beta)$.
In an integrable theory, the $S$ matrix must satisfy the Yang-Baxter
equation. This ensures that the multi-particle scattering
amplitudes factorize into the sum of products of two-particle amplitudes,
a requirement of integrability. The simplest solution
of these constraints is  \cite{ZZ}
\begin{eqnarray}
\nonumber
\frac{S_1(\beta)}{S_0(\beta)}&=&\frac{2\pi i}{(N-2)\beta}\\
\label{onsmat1}
\frac{S_2(\beta)}{S_0(\beta)}&=&\frac{2\pi i}{(N-2)(i\pi -\beta)}\\
\nonumber
S_0(\beta) &=& -\frac{
\Gamma\left(1+\frac{i\beta}{2\pi}\right)
\Gamma\left(\frac{1}{2} -\frac{i\beta}{2\pi}\right)
\Gamma\left(\frac{1}{N-2} -\frac{i\beta}{2\pi}\right)
\Gamma\left(\frac{1}{2}+\frac{1}{N-2} +\frac{i\beta}{2\pi}\right)}
{\Gamma\left(-\frac{i\beta}{2\pi}\right)
\Gamma\left(\frac{1}{2} +\frac{i\beta}{2\pi}\right)
\Gamma\left(1+\frac{1}{N-2} +\frac{i\beta}{2\pi}\right)
\Gamma\left(\frac{1}{2}+\frac{1}{N-2} -\frac{i\beta}{2\pi}\right)}
\end{eqnarray}

A number of checks indicate that the entire spectrum
of the sigma model consists of these $N$ particles
with this $S$ matrix.  It agrees with large-$N$ computations
\cite{ZZ}. By computing the energy at zero temperature in
the presence of a magnetic field, one can obtain correctly the first
terms of the beta function \cite{Hasen}.  Another substantial check is
to compute the free energy of this massive theory at non-zero
temperature from this $S$ matrix.  The free energy in the
weak-coupling limit must agree with that of $N-1$ free bosons (the
action (\ref{onaction}) as $g\to 0$), and indeed does
\cite{mesigmacoset}. These checks effectively prove
the ansatz that (\ref{onsmat}, \ref{onsmat1}) are the
exact $S$ matrix for the $O(N)$ sigma model for $N>2$.

This $S$ matrix involves $N$ particles, and so seems to require that
$N$ needs to be an integer to make sense. I will discuss below how to
define an $S$ matrix for continuous $N$, but 
to see what happens as $N\to 2$, it is useful to examine
the zero-temperature
energy in a magnetic field $H$. This is a standard computation in an
integrable model with a $U(1)$ symmetry. At zero temperature, particles
with rapidity $|\beta|<B$ fill a Fermi sea, where $B$ depends on $H$.
The energy of this sea is
\begin{equation}
E(H)-E(0)= -{m\over 2\pi}\int_{-B}^{B}d\beta\ \cosh\beta
\ \epsilon(\beta)
\label{forE}
\end{equation}
where $\epsilon(\beta)$ is the energy lost by removing a particle
from the sea. It is given by the equation
\begin{equation}
\epsilon(\beta)=H - m\cosh \beta  +
\int_{-B}^{B} d\beta' \kappa(\beta -\beta')\epsilon(\beta').
\label{forep}
\end{equation}
where 
$\kappa(\beta)$ follows from the $S$ matrix, and for this model is 
\cite{Hasen}
\begin{eqnarray}
\kappa^{O(N)}(\beta) &\equiv&
\frac{1}{2\pi i}\frac{d}{d\beta}\ln\left[S_0(\beta)+S_1(\beta)\right]\\
&=&\int_{-\infty}^\infty \frac{d\omega}{2\pi} e^{i\omega\beta} 
e^{-\pi|\omega|/(N-2)}\frac{\cosh\left(\frac{N\pi}{2(N-2)}|\omega|\right)}
{\cosh\left(\frac{\pi}{2}\omega\right)}
\label{onkernel}
\end{eqnarray}
The value of $B$ is determined by imposing the condition $\epsilon(B)=0$.
Clearly these equations are valid and can be solved for continuous $N\ge 2$.

One interesting thing to note is that the above $S$ matrix and energy
do make sense right at $N=2$. Moreover, it correctly describes the
continuum limit of the lattice model (\ref{poly},\ref{loopgas}) 
when $N=2$.  This
is straightforward to check: when $N=2$, the lattice model is equivalent to
the classical XY rotor right at its Kosterlitz-Thouless point
\cite{nienhuis}. The appropriate field theory description is indeed a
single boson perturbed by a marginally-relevant operator (the vortex
creation operator).  Explicitly, this field theory
is the sine-Gordon
model with action
\begin{equation}
S^{SG}=\int d^2 x\  \left[\frac{1}{2}\partial_\mu \varphi \partial^\mu
\varphi + G\cos(\sqrt{8\pi}\varphi)\right].
\label{sg}
\end{equation} 
Here the role of the coupling $g$ is played by the coefficient $G$ of
the $\cos\beta_{SG}\varphi$ term.  This point $\beta_{SG}^2 =8\pi$ is
where the sine-Gordon model closely resembles a sigma model: not only
is the operator perturbing the trivial fixed point marginally
relevant, but the model has a Lie-algebra symmetry.  In fact, it turns
out to have a symmetry larger than the $O(2)$ it satisfies by
construction: the two particles form a doublet under a larger $SU(2)$
symmetry. In the sine-Gordon language, the particle doublet consists
of a kink and an antikink in the field $\varphi$; there are no other
particles in the spectrum at this value of $\beta_{SG}$. In
particular, there is no single-particle state corresponding to the
field $\varphi$ itself: $\varphi$ creates only multi-particle states
with the same number of kink as antikinks.  Note that the action
(\ref{onaction}) with $N$=2 is {\bf not} sufficient to describe this
field theory: defining $\phi_1=\cos[\sqrt{g/2}\varphi]$,
$\phi_2=\sin[\sqrt{g/2}\varphi]$ reproduces the first but not the
second term in (\ref{sg}).  For $N>2$, one argues that
(\ref{onaction}) holds because there are no relevant terms one can add
consistent with the $O(N)$ symmetry. However, as seen from (\ref{sg}),
there indeed is such a term when $N=2$.  In this sense, the $S$ matrix
description is more closely related to the original non-perturbative
lattice formulation than is the sigma model. In fact, note that this
$S$ matrix at $N=2$ describes the physics near the {\it non-trivial}
critical point at $K=K_c=1/\sqrt{2}$, not the trivial fixed point at
$K=\infty$.  Thus continuing the $N> 2$ $S$ matrix to $N=2$ does give
a sign of the new physics occurring for $N<2$.

Although the $S$ matrix is valid at $N=2$, it cannot be continued
past: a singularity at $N=2$ is immediately apparent from the explicit
form of the above $S$ matrix. Such a singularity appears in all
integrable $O(N)$-invariant field theories in two dimensions.
Continuing around the singularity leads to nonsensical results. For
example, the kernel $\kappa^{O(N)}(\beta)$ makes no sense for $N<2$,
because the integral does not converge for real $\beta$.
Since a naive continuation of the $N<2$ $S$ matrix to $N<2$ is
not possible, one must therefore formulate a
different ansatz.  Such an ansatz was given
in \cite{Zpoly}; I will present a slightly different argument for
this result here. 
Based on what is known for $N\ge 2$, the following criteria for such
an $N\le 2$ $S$ matrix seem reasonable:
\begin{enumerate}
\item The particles should be in the vector representation of $O(N)$.
\item The $S$ matrix as $N\to 2$ from below should be the same
as $N\to 2$ from above.
\end{enumerate}
Note that the second criterion does not rule out a singularity at
$N=2$: derivatives of the $S$ matrix elements with respect to $N$ need
not be continuous.

These criteria are sufficient to infer the $N<2$ $S$ matrix.
The first criterion means that the $S$ matrix must still be of the form
(\ref{onsmat}). However, we must look for a solution of the Yang-Baxter
equation other than (\ref{onsmat1}). Such a solution has long been
known to exist \cite{oldon,Zpoly}. It is
\begin{eqnarray}
S_0 (\beta)&=& 0\\ 
\frac{S_2 (\beta)}{S_1(\beta)} 
&=& \frac{\sinh\left[\mu(i\pi -\beta)\right]}
{\sinh\left[\mu\beta\right]}
\end{eqnarray}
where $\mu$ is defined by
$$N= 2\cos\left({\pi\mu}\right).$$
Crossing and unitarity yields 
\begin{eqnarray*}
S_1 (\beta) &=& \frac{\Gamma\left(\frac{1}{2} - \frac{\mu\beta}{i\pi}\right)}
{\Gamma\left(\frac{1}{2} + \frac{\mu\beta}{i\pi}\right)}
\prod_{k=1}^\infty \frac{
\Gamma\left(2k\mu - \frac{\mu\beta}{i\pi}\right)
\Gamma\left(1+2k\mu - \frac{\mu\beta}{i\pi}\right)}
{\Gamma\left(2k\mu + \frac{\mu\beta}{i\pi}\right)
\Gamma\left(1+2k\mu + \frac{\mu\beta}{i\pi}\right)}\\
&&\qquad\qquad\times\frac{
\Gamma\left((2k-1)\mu + \frac{\mu\beta}{i\pi}\right)
\Gamma\left(1+(2k-1)\mu +\frac{\mu\beta}{i\pi}\right)}
{\Gamma\left((2k-1)\mu - \frac{\mu\beta}{i\pi}\right)
\Gamma\left(1+(2k-1)\mu - \frac{\mu\beta}{i\pi}\right)}
\end{eqnarray*}
This $S$ matrix makes sense for $N\le 2$; in fact it does not make
sense when $N>2$ because $\mu$ becomes imaginary.  It agrees with the
$N\ge 2$ $S$ matrix when $N=2$.  It also has a nice intuitive
interpretation: because $S_0$ vanishes, the world lines of the
particles do not cross and one can think of them as the loops in the
loop gas (\ref{loopgas}).  These points led Zamolodchikov to
conjecture that it describes the $O(N)$ lattice model for $N<2$
\cite{Zpoly} at the dilute/dense critical point and in the dilute
phase.

This conjecture was proven in \cite{smirnov}. A key element of the
proof is that when one inserts the form (\ref{onsmat}) into the
Yang-Baxter equation, the equation reduces to a set of functional
relations relating $S_0$, $S_1$ and $S_2$. These relations involve $N$
only as a parameter, so the functions can be found for any $N$. This
suggests there is an $S$ matrix with identical physical properties
valid at any $N$ \cite{smirnov}.  To find this equivalent $S$ matrix,
note that when $S_0(\beta)=0$, the $S$ matrix is
\begin{eqnarray*}
S = S_1(\beta) P + S_2(\beta) e,
\end{eqnarray*}
where $P$ and $e$ are $N^2\times N^2$ matrices. However, to compute
quantities like the free energy, the explicit matrix form of $P$ and
$e$ is not required. All that needs to be known is what algebra the
matrices obey (for example, $P$ is the permutation operator, and $e^2
= N\, e$). In a different matrix representation of the same algebra,
all physical consequences will be the same. In the polymer case, the
algebra is called a Temperley-Lieb algebra, and representations
exist for any $N<2$ \cite{smirnov}.  

The $S$ matrix approach therefore allows a precise {\it continuum}
formulation of the model valid for any $N$ on and off the critical
point.  Moreover, the critical limit ends up being equivalent to the
Coulomb-gas description, which in the continuum is a free boson with a
charge at infinity \cite{DF}.  This proof means that all $S$ matrix
computations give identical results to those coming from the continuum
Coulomb-gas description of the $O(N)$ model.  In fact, since the $S$
matrix description is valid non-perturbatively, one can compute new
results such as as the scaling function for the number of polymers on
a cylinder \cite{FSpoly} as well as correlators from
the form-factor expansion \cite{CM}.

The $S$ matrix derived in \cite{Zpoly} thus gives a natural extension
of the $O(N)/O(N-1)$ sigma model to $N<2$. It is motivated by the loop
gas, but the above criteria for finding the $N<2$ $S$ matrix arise
from studying only the $N>2$ $S$ matrix, not the lattice model. It
thus gives hope that other $O(N)$-invariant sigma models defined for
$N>2$ can be extended to $N<2$.

\section{Speculations on the GSE class} 

Unfortunately, what seems so clear in the polymer case is very muddy
in the GSE case. I do not know of a lattice model in the same
universality class. I also do not know of an $S$ matrix for $P<1$.  In
this section I discuss the situation, and one $S$ matrix which seems
plausible but which does not yield the phase diagram of section 2.

The massive $O(2P)/O(P)\times O(P)$ field theory at $\theta=0$ and the
massless one at $\theta=\pi$ are both integrable, and their exact
spectrum and $S$ matrices are known for $P>1$ \cite{metheta,meTBA}.  A
crucial fact is that both have kink states, and that at $P=1$ and
$P=2$ these are the only states which remain in the spectrum.  Thus to
understand what happens for $P<1$, one needs to concentrate on the
kinks.  The kinks in the $O(2P)/O(P)\times O(P)$ sigma model at
$\theta=\pi$ are in the two spinor representations of $O(2P)$
\cite{metheta}, each of dimension $2^{P-1}$.  At $\theta=0$, they are
in the representations with highest weight $2\mu_s$ and
$2\mu_{\overline s}$ (where $\mu_s$ and $\mu_{\overline s}$ are the
highest weights of the spinor representations). These can be obtained
by taking the symmetric tensor product of two $s$ and two $\overline
s$ representations respectively, and each is of dimension
$(2P-1)!/P!(P-1)!$ \cite{metheta,meTBA}.

The complete explicit form of the kink $S$ matrix for $P\ge 1$
is quite
complicated and not particularly illuminating. It does enable one
to find the kernel $\kappa(\beta)$ in the equations for the energy
in a background magnetic field, as described in
(\ref{forE},\ref{forep}). The magnetic field can be chosen
so that the Fermi sea is filled with only
one kind of kink, corresponding to the state with weight equal
to the highest weight $2\mu_s$. The resulting kernel
for the sigma model at $\theta=0$ is
\begin{equation}
\kappa^{sigma}(\beta) 
=\int_{-\infty}^\infty \frac{d\omega}{2\pi} e^{i\omega\beta} 
\left[1-e^{-\pi|\omega|/(2P-2)}
\frac{\sinh\left(\frac{P\pi}{(2P-2)}|\omega|\right)}
{\cosh\left(\frac{\pi}{2}\omega\right) }
\right].
\label{sigmakernel}
\end{equation}
This kernel cannot be continued  naively to $P<1$.
For the sigma model at $\theta=\pi$, the equations (\ref{forE})
and (\ref{forep}) are modified
because the particles are massless, but as described in detail
in \cite{metheta}, the most important kernel is the same as at $\theta=0$.

As with the polymers, it is important to understand what happens at
the singularity, here $P=1$. Unlike the polymers, there is no known
underlying lattice model to appeal to for guidance. However, the
polymer case taught a valuable lesson in that the $N>2$ $S$ matrix was
still valid for $N=2$. Even more remarkably, it described the physics
in the region around the non-trivial fixed point which just appeared
at $N=2$. This of course may have just been a stroke of luck peculiar
to that problem. But since the action of the sigma model at $P=1$
reduces to that of a free boson, the $S$ matrix is the main piece of
information potentially useful here. Thus to proceed it is necessary
to assume that it is meaningful to continue
the spectrum and $S$ matrix for $P>1$ to $P=1$.

The first thing to note is that for $P=1$ the dimensions of the
representations with highest weights $\mu_s, \mu_{\overline s}, 2\mu_s$
and $2\mu_{\overline s}$ are each $1$. Thus when $P=1$
there should be two
particles. The $P\to 1$ limit of (\ref{sigmakernel}) is
equal to the $N\to 2$ limit of (\ref{onkernel}). Thus
$$S_{ss}=S_{1}+S_2 = 
\frac{\Gamma\left(1-\frac{\beta}{2\pi i}\right)
\Gamma\left(\frac{1}{2}+\frac{\beta}{2\pi i} \right)}
{\Gamma\left(1+\frac{\beta}{2\pi i}\right)
\Gamma\left(\frac{1}{2}-\frac{\beta}{2\pi i}\right)}
$$ here.  For $N=2$ in the $O(N)$ model,
backscattering between the two particles is allowed, because they came
from the same multiplet in the vector of $O(N)$. For $P>1$ in
the sigma model,
backscattering between $s$ particles and $\overline{s}$ is forbidden,
but the extra conservation law requiring this reduces to ordinary
energy conservation when $P=1$.  Thus when $P=1$, 
the one $s$ particle and the one $\overline{s}$ can backscatter.
The correspondence of this $S$ matrix element and the possibility of
backscattering means that it is plausible that the sigma model
at $\theta=0$ is identical to the $N=2$ $O(N)$ model,
namely the sine-Gordon model (\ref{sg}) at $\beta_{SG}^2 = 8\pi$.

Now comes the speculating. 
To proceed, one needs to impose criteria
for a $P<1$ $S$ matrix. A reasonable set of criteria for $\theta=0$ is
\begin{enumerate}
\item The massive particles for $P<1$ are in the 
representations of $O(2P)$ with highest weights 
$2\mu_s$ and $2\mu_{\overline s}$.
\item Backscattering between the two representations occurs.
\item As $P\to 1$, the $S$ matrix is that of
the sine-Gordon model at $\beta_{SG}^2=8\pi$, which happens
to be $SU(2)$ symmetric.
\end{enumerate}
\medskip
For $\theta=\pi$, the representations are instead the spinors
$s$ and $\overline{s}$.

The first criterion means that the $S$ matrix describes the scattering
of $M$ particles, where $M=2(2P-1)!/P!(P-1)!$ for $\theta=0$, and
$M=2^P$ for $\theta=\pi$.  The third means that for $P=1$, the model
is $SU(2)$ symmetric.  The second criterion suggests that we group the
particles in the $s$ and $\overline s$ representations into one
representation of some larger symmetry. Since there are $M$ particles,
and $M=2$ when $P=1$, the simplest possibility seems to be that these
particles are in the {\it vector} representation of $SU(M)$.
Therefore an $S$ matrix with particles in the vector of
$SU(M)$ satisfies all three criteria. Such an $S$ matrix satisfying
all the constraints of integrability is well known, and describes
scattering in the $SU(M)$ ``chiral'' Gross-Neveu model \cite{sun}. The
critical point in this model is the $SU(M)_1$ conformal field theory.
If this $S$ matrix is the correct continuation to $P<1$, then the
metal-insulator transition is described by the conformal field theory
$\lim_{M\to 1} SU(M)_1.$ A simple but substantial check on the truth
of this conjectured $S$ matrix is to compute the central charge $c$ of
the conformal field theory describing the critical point. If the
conformal field theory is to describe a phase transition in a
disordered system, it must have $c=0$. The central charge of $SU(M)_1$
is $c = M-1$, so as $P\to 0$, $c\to 0$ as required.  This is not an
immediate consequence of the criteria imposed, so I view this as a
good indication that the whole procedure is reasonable.

Unfortunately, this $S$ matrix and conformal field theory do not seem
to describe the physics of the 2d GSE metal-insulator transition. 
This is apparent from the behavior of the
model away from the critical point. In the Gross-Neveu model, the
operator perturbing the model away from the critical point is of
dimension 2 (naively marginal). Computing the beta function to first
non-trivial order yields that the operator is marginally relevant with
one sign of coupling constant, and marginally irrelevant with the
other sign. This contradicts the phase diagram in the right half of
figure 1: at the non-trivial critical point the perturbing operator is
relevant in both directions, not just one. Moreover, a perturbing
operator of dimension 2 implies that the thermal exponent
$\nu=\infty$.  This is possible, but is not very consistent with the
most recent numerical results, which suggest that $\nu$ is around
$2.5$ \cite{numerical}. 

This is why I do not know what the appropriate $S$ matrix is for this
class of model, or for that matter, if one even exists.  A key missing
ingredient is a lattice model in the same universality class.  The
most likely candidate is a generalization of (\ref{poly}) with
spins in the symmetric representation of $O(2N)$ instead of the
vector, but I have not been able to develop a Coulomb-gas
formulation for such a model. If this were to be accomplished,
this would shed a great deal of light on the problem.  Another
interesting avenue to explore would be to understand for a given
disordered system whether $\theta=0$ or $\theta=\pi$. It seems likely
that $\theta$ here is not a tunable parameter, like the coefficient of
the Wess-Zumino-Witten term appearing in other sigma models for
disordered systems \cite{FK}.

I should also note that an analogous continuation of the $O(N)$
Gross-Neveu model to $N<2$ was proposed in \cite{giuseppe}. This
proposal involves particles in the vector representation of
$O(N)$. Since particles in the vector representation of $O(N)$ do not
appear in the Gross-Neveu model for $N=3$ or $4$, this proposal does
not really obey the sort of criteria discussed in this paper.
Nevertheless, there does not exist any better proposal yet, so the
definitive answer for this situation as well as the GSE class remains
unknown.

\bigskip\bigskip I would like to thank D.~Bernard and N.~Read for
conversations, and in particular for convincing me that the $S$ matrix
discussed at the end does not apply to the GSE class.  My work is
supported by by NSF grant DMR-0104799. a DOE OJI Award, and a Sloan
Foundation Fellowship.

\end{document}